\documentclass[rnote]{aa}
\usepackage{graphicx}
\usepackage{txfonts}
\newcommand{\rg}{R_{\mbox{\tiny{S}}}}
\newcommand{\pder}[2]{\frac{\partial #1}{\partial #2}}
\newcommand{\U}{\mathcal{U}}
\newcommand{\ratio}[2]{{\textstyle \frac{\,#1}{\,#2}}}
\newcommand{\D}[1]{D_{#1}}
\newcommand{\A}{\widehat{A}}
\newcommand{\Kappa}{\mathrm{K}}
\newcommand{\ii}{\mathrm{i}}
\newcommand{\dd}{\mathrm{d}}
\newcommand{\const}{\mbox{const}}

\begin{document}
\title{Twin-peak quasiperiodic oscillations as an internal resonance}
\author{J.~Hor\'ak \and V.~Karas}
\institute{Astronomical Institute, Academy of Sciences, Bo\v{c}n\'{\i}~II, CZ-141\,31~Prague, Czech Republic}

\authorrunning{J.~Hor\'ak \& V.~Karas}
\titlerunning{Twin-peak quasiperiodic oscillations as an internal resonance}
\date{Received 12 August 2005; accepted 30 December 2005}
\abstract{}{Two inter-related peaks occur in high-frequency power
spectra of X-ray lightcurves of several black-hole candidates. We
further explore the idea that a non-linear resonance mechanism,
operating in strong-gravity regime, is responsible for these
quasi-periodic oscillations (QPOs).}{By extending the multiple-scales
analysis of Rebusco, we construct two-dimensional phase-space sections,
which enable us to identify different topologies governing the system
and to follow evolutionary tracks of the twin peaks. This
suggests that the original (Abramowicz \& Klu\'zniak)
parametric-resonance scheme can be viewed as an ingenuous account of
the QPOs model with an internal resonance.}{We show an example of
internal resonance in a system with up to two critical points, and we
describe a general technique that permits to treat other cases in a
systematical manner. A separatrix divides the phase-space sections into
regions of different topology: inside the libration region the
evolutionary tracks bring the observed twin-peak frequencies to an exact
rational ratio, whereas in the circulation region the observed
frequencies remain off resonance. Our scheme predicts the power should
cyclically be exchanged between the two oscillations. Likewise the
high-frequency QPOs in neutron-star binaries, also in black-hole sources
one expects, as a general property of the non-linear model, that slight
detuning pushes the twin-peak frequencies out of sharp resonance.}{}
\maketitle

\section{Introduction}  
Twin peaks occur in high-frequency ($\sim50$--$450$~Hz) power spectra of
X-ray ($\sim2$--$60$~keV) lightcurves of several black-hole candidates (see
van der Klis \cite{kli06}; McClintock \& Remillard \cite{mcc06} for
recent reviews of observational properties and theoretical
interpretations). This transient phenomenon seems to be connected with
the kilohertz quasi-periodic oscillations (QPOs) in neutron star
sources, of which more examples are known (about the tens at present).
The nature of black hole high-frequency QPOs remains puzzling despite
variety of models proposed in the literature. In neutron-star low-mass
X-ray binaries these twin QPOs are known to occur often simultaneously
and they can be highly coherent ($Q\gtrsim10^2$; e.g.\ Barret et al.\
\cite{bar05}), slowly wandering in frequencies between different
observations, whereas in black-hole candidates the QPO coherency appears
to be lower ($Q\sim2$--$10$) and the presence of a pair pops up only
when a collection of observations is carefully analyzed. 

In Abramowicz \& Klu\'zniak (\cite{abr01}) and Klu\'zniak \& Abramowicz
(\cite{klu01}) an idea of accretion disc resonance was proposed, which
naturally incorporates pairs of frequencies occurring in a ratio of
small integer numbers. This scheme  predicts
the observed frequency ratios in black-hole QPO sources should prefer
the $3$:$2$ ratio; it also  suggests this could be understood if a
non-linear coupling mechanism operates in a black-hole accretion disc,
where strong-gravity effects are essential. Indeed, especially in those
black-hole candidates where the high-frequency QPOs have been reported,
they occur very close to the ratio of small integer numbers, $3$:$2$ in
particular (Miller et al.\ \cite{mil01}; Strohmayer \cite{str01}; Homan
et al.\ \cite{hom05}; Remillard et al.\ \cite{rem05}; Maccarone \&
Schnittman \cite{mac05}). Nowadays, the original account can be viewed
as a na\"{\i}ve model with the internal resonance. Various realizations 
of this scheme have been
examined in terms of accretion disc/torus oscillations (e.g.\ Abramowicz
et al.\ \cite{abr03}; Bursa et al.\ \cite{bur04}; Kato \cite{kat04}; Li
\& Narayan \cite{li_04}; Schnittman \& Rezzolla \cite{sch05}; Zanotti et
al.\ \cite{zan05}).

So far the ``right'' model has not yet been identified. However, it has 
been recognized that fruitful knowledge about common properties of
high-frequency QPOs can be gained by investigating a very general
resonance scheme, which likely governs matter near a compact accreting
body. To this aim, Abramowicz et al.\ (\cite{abr03}) examined the 
epicyclic resonances in a nearly-geodesic motion in strong gravity.
Rebusco (\cite{reb04}) and Hor\'ak (\cite{hor04}), by employing the
method of multiple scales (Nayfeh \& Mook \cite{nay79}), have
demonstrated that the $3$:$2$ resonance is indeed the most prominent 
one near horizon of a central black hole. Only certain frequency
combinations are allowed, depending on symmetries which the system
exhibits, and only some of the allowed combinations have chance to give
rise to a strong resonance.

In the present paper we pursue this approach further and we find tracks
that an axially symmetric system with two degrees of freedom, near
resonance, should follow in the plane of energy (of the oscillations)
versus radius (where the oscillations take place). We show different
topologies of the phase space in the way that closely resembles the
method of disturbing function, familiar from the studies of the
evolution of mean orbital elements in celestial mechanics (Kozai
\cite{koz62}; Lidov \cite{lid62}). The analogy is very illuminating
and it provides a systematic way of distinguishing topologically
different states of the system. In particular, one can discriminate
regions of phase space where the observed frequency ratio fluctuates
around an exact rational number from those regions where this ratio
remains always outside the resonance. Our model suggests that even
black-hole twin QPOs should vary in frequency and they should not stay
at a firmly fixed frequency ratio, albeit the expected variation is very
small -- certainly less than what has been frequently reported in neutron 
star binaries and what can be tested with the data available at present.

\section{A conservative system with two degrees of freedom}
\subsection{Non-linear terms in the governing equations}
\label{sec:epc-ring}
Let us consider an oscillatory system with two degrees of freedom, which
is described by coupled differential equations of the form
\begin{eqnarray}
\label{eq:res_gov_r}
\ddot{\delta\rho} + \omega_r^2\; \delta\rho &=&
  f_\rho(\delta\rho,\delta\theta, \dot{\delta\rho},\dot{\delta \theta}),
\label{eq:ms-generalfrh}
  \\
\label{eq:res_gov_theta}
\ddot{\delta \theta} + \omega_\theta^2\; \delta \theta &=&
  f_\theta(\delta\rho, \delta\theta, \dot{\delta\rho}, \dot{\delta \theta}).
\label{eq:ms-generalfth}
\end{eqnarray}
We assume that the right-hand-side functions are nonlinear (their 
Taylor expansions start with the second order), and that they are
invariant under reflection of time. Clearly, these equations include the
case of a nearly circular motion under the influence of a perturbing
force: $\delta\rho$ and $\delta\theta$ are small perturbations of the
position, whereas $\omega_r(r)$ and $\omega_\theta(r)$ have a meaning of
the radial and the vertical epicyclic frequencies along the circular
orbit $r=r_0$, $\theta=\pi/2$:
\begin{equation}
 \omega_r^2=\frac{\partial^2\U}{\partial r^2},
 \quad
 \omega_\theta^2=\frac{1}{r_0^2}
 \left(\frac{\partial^2\U}{\partial\theta^2}\right),
\label{eq:epc-om}
\end{equation}
where the effective potential is
\begin{equation}
 \U(r,\theta)\equiv\Phi(r,\theta)+\frac{\ell^2}{2r^2\sin^2\theta},
\label{eq:epc-defU}
\end{equation}
$\Phi(r,\theta)$ is the gravitational potential, for which axial
symmetry and staticity will be assumed. These assumptions make our
system qualitatively different from models requiring non-axisymmetric
perturbations.

\begin{figure*}[t]
\includegraphics[width=0.49\textwidth]{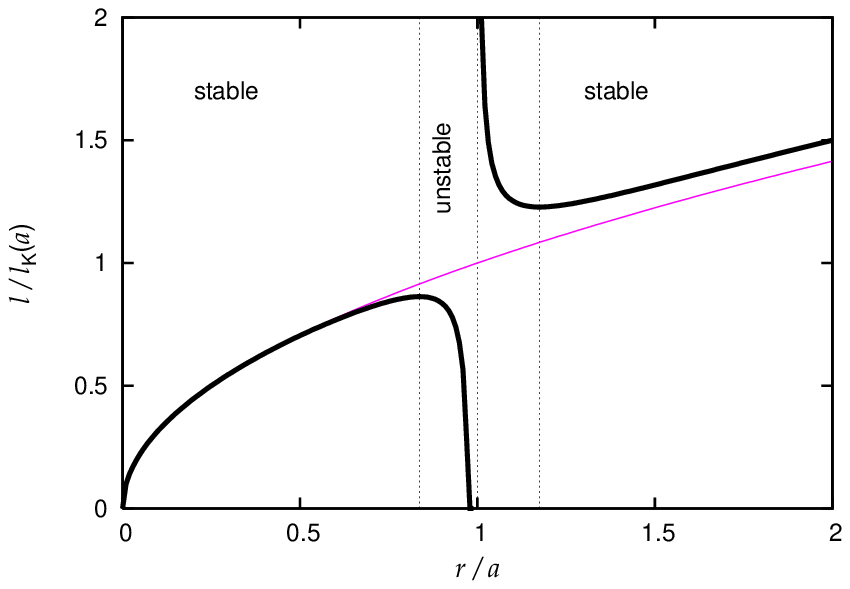}
\hfill
\includegraphics[width=0.49\textwidth]{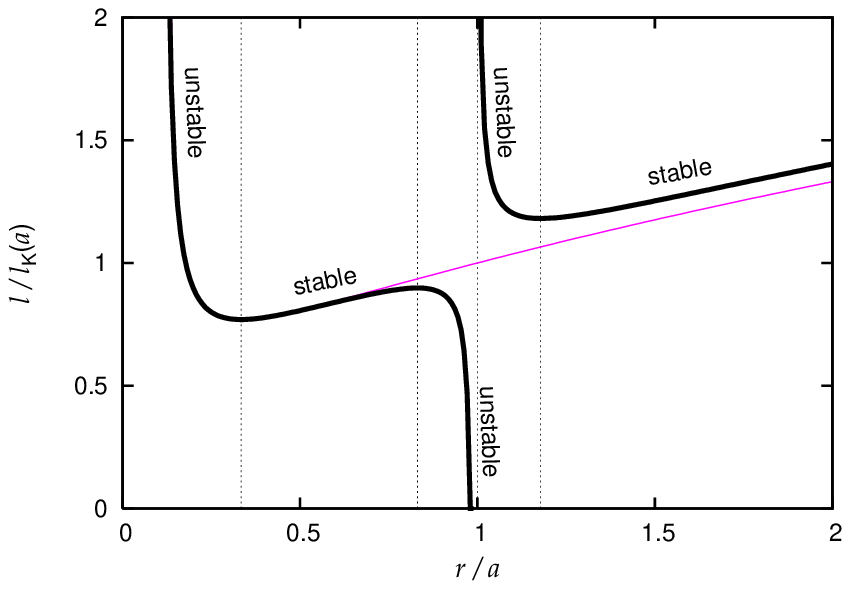}
\caption{Angular momentum $\ell(r)$ (thick line) of a test particle on a
circular orbit in the combined gravitational field of a spherical body
and a ring. The mass of the ring relative to the central mass is
$\mu=0.1$ and the particle mass is set to unity. Left panel: the case of
Newtonian central field. Right panel: the pseudo-Newtonian case. Radius
has been scaled with respect to the ring radius (here, $a=9\rg$), and the
angular momentum has been scaled by the value of Keplerian angular
momentum $\ell_\mathrm{K}(a)$. Keplerian angular momentum in the
central field is also plotted (thin line). Circular orbits are Rayleigh
unstable and the epicyclic approximation is inadequate in regions where
the angular momentum decreases with radius.}
\label{fig:ring-l}
\end{figure*}

\begin{figure*}[t]
\includegraphics[width=0.49\textwidth]{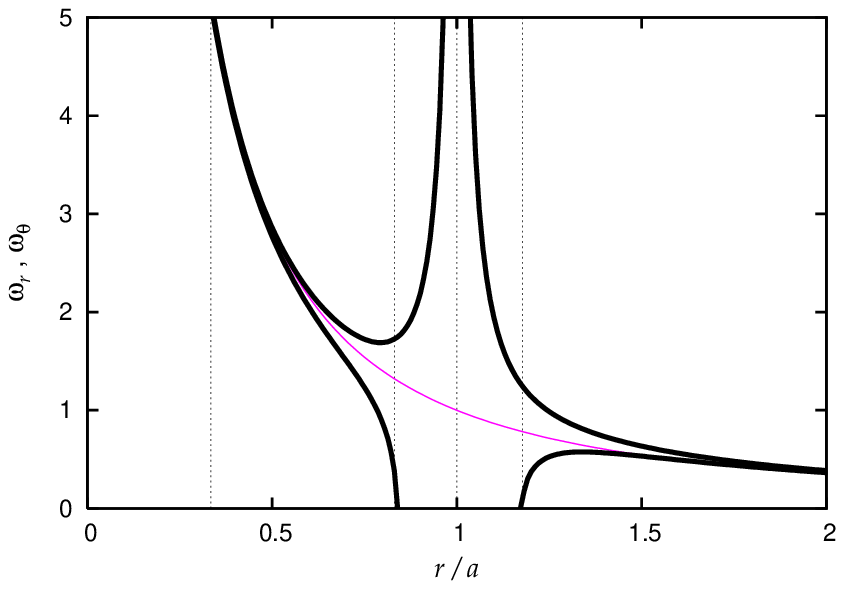}
\hfill
\includegraphics[width=0.49\textwidth]{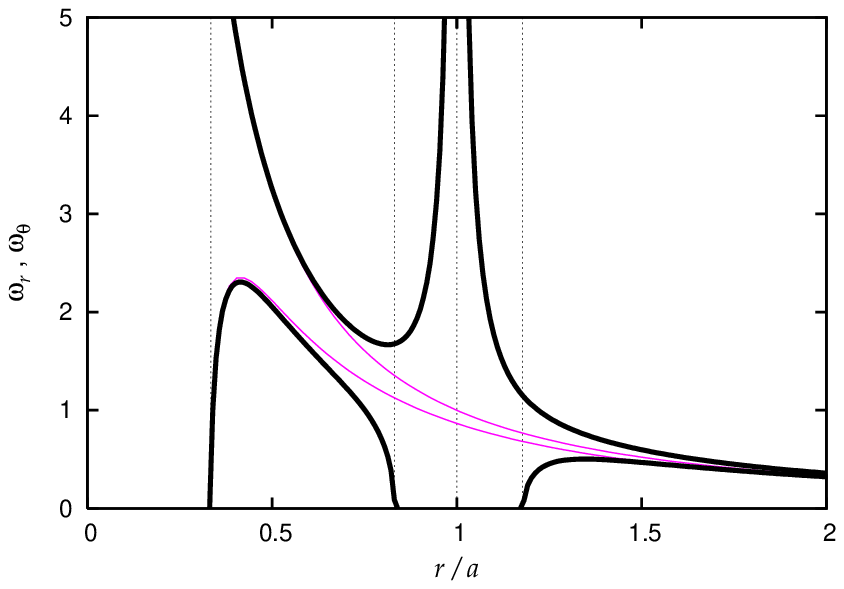}
\caption{The behaviour of epicyclic frequencies (thick lines;
$\omega_\theta$ -- top, $\omega_r$ -- bottom) provides basic clues to
the origin of oscillations and the effects expected to occur in strong
gravity. Left/right panels refer to the Newtonian/pseudo-Newtonian
cases, as in the previous figure. Units of the ordinates are scaled by
$\sqrt{GM/a^3}$ to make frequencies dimensionless. The frequencies in
absence of the ring ($\mu=0$) are also indicated for reference (thin
lines; $\omega_\theta(r)=\omega_r(r)$ in the Newtonian case).}
\label{fig:ring-w}
\end{figure*}

As a generic example, let the total gravitational field be given as a
superposition of the central potential of a spherical star,
$\Phi_\mathrm{s}(r)$, plus an axisymmetric term
$\Phi_\mathrm{r}(r,\theta)$,
\begin{equation}
 \Phi(r,\theta)=\Phi_\mathrm{s}(r)+\Phi_\mathrm{r}(r,\theta).
\label{eq:potential}
\end{equation}
For the central field we assume the form
\begin{equation}
 \Phi_\mathrm{s}(r)=-\frac{GM}{\tilde{r}},
\label{eq:epc-starPhi}
\end{equation}
where we set $\tilde{r}=r$ or $\tilde{r}=r-\rg$ (to adopt the Newtonian
or the pseudo-Newtonian approximations; $\rg\equiv2GM/c^2$). We assume a
circular ring (mass $m$, radius $a$) as a source of the perturbing
potential, 
\begin{equation}
 \Phi_\mathrm{r}(r,\theta)=-\frac{2Gm}{\pi}\frac{K(k)}{B^{1/2}},
\label{eq:epc-ringPhi}
\end{equation} where $K(k)$ is a complete elliptical integral of the
first kind, $B(r,\theta)\,\equiv\,r^2+a^2+2ar\sin\theta$,
$k(r,\theta)\,\equiv4\,ar\sin\theta/B(r)$. 

At this point a remark
is necessary regarding the interpretation of the potential 
(\ref{eq:potential})--(\ref{eq:epc-ringPhi}): we conceive it
as a toy-model for strong-gravity effects and internal resonances that we seek in
the system of a black hole and an accretion disc, but the origin of the
perturbing potential $\Phi_\mathrm{r}(r,\theta)$ is not supposed to be 
the gravitational field of the accretion disc itself. Of course the inner 
disc is not self-gravitating in black hole binaries and the ring
is not introduced here with the aim of representing the accretion flow
gravity. What we imagine is that hydrodynamic and magnetic forces 
are producing qualitative effects, which can be captured by the ring potential
in our equations (\ref{eq:ms-generalfrh})--(\ref{eq:ms-generalfth}).
On the other hand, our approach is rather general and it is worth remembering 
that the same formalism can be successfully applied also in systems where 
the disc self-gravity plays a non-negligible role.

Also in general relativity, weakly perturbed (i.e.\
nearly-geodesic) motion of gas elements orbiting around a Schwarzschild
black hole can be described by effective potential that contains a
spherical term arising from the gravitational field of the central black
hole, plus a perturbing term, which we assume axially symmetric.
Naturally, the interpretation of the perturbing term is more complicated
if one would like to derive its particular form from an exact solution
of Einstein's equations. We do not want to enter into complications of a
special model here (but see Letelier \cite{let03}; Karas et al.\
\cite{kar04}; Semer\'ak \cite{sem04} and references cited therein  for a
review and examples of spacetimes that contain a black hole and a
gravitating ring in general relativity). Also, one may ask whether the
adopted approach can comprehend frame-dragging effects of Kerr spacetime
while the black hole rotation is considered as perturbation to the
spherical field of a static non-rotating black hole; even if a general
answer to that question would be affirmative, the Kerr metric has rather
special properties concerning the integrability of the geodesic motion
and the form of non-sphericity, which quickly decays with radius. Our
current opinion is such that the rotation-related effects of Kerr metric
cannot be viewed as the origin of the perturbation required for
black-hole high-frequency twin QPOs.

The angular momentum of a test particle orbiting the centre on an
equatorial circular orbit is
\begin{equation}
 \ell{\equiv}r^3\left(\pder{\Phi}{r}\right)= 
 GMr\left[\frac{r^2}{\tilde{r}^2} +
 \mu r\frac{(r+a)E(k_0)+(r-a)K(k_0)}{\pi(r^2-a^2)}\right],
\end{equation}
where $E(k)$ is a complete elliptical integrals of the second kind and the
right-hand-side terms are evaluated at the orbit radius,
$k_0{\equiv}k(r_0,\pi/2)$, $\mu{\equiv}m/M$. Figure~\ref{fig:ring-l}
captures a typical curve of the angular momentum as a function of
radius. The corresponding orbital frequency is
\begin{equation}
 \Omega^2(r) = \frac{GM}{r^3}\left[\,\frac{r^2}{\tilde{r}^2}+
 \mu r\frac{(r+a)E(k_0)+(r-a)K(k_0)}{\pi(r^2-a^2)}\right].
 \label{eq:epc-ringOm}
\end{equation}
Equation~(\ref{eq:epc-om}) gives the epicyclic frequencies,
\begin{eqnarray}
 \omega_r^2(r) \!\!&=&\!\!
 \frac{GM}{r^3}\left[\,\frac{r^2(3\tilde{r}-2r)}{\tilde{r}^3}+\frac{2\mu}{\pi}\,
 \frac{(r-a)^2 K(k_0)-a^2 E(k_0)}{r^2(r-a)^2(r+a)}\right],
\label{eq:epc-ringomr}\\
 \omega_\theta^2(r) \!\!&=&\!\!
 \frac{GM}{r^3}\left[\,\frac{r^2}{\tilde{r}^2}+\frac{2\mu}{\pi}\,
 \frac{E(k_0)}{(r-a)^2(r+a)}\right].
\label{eq:eoc-ringomth}
\end{eqnarray}
The epicyclic frequencies as functions of radius are plotted in
Figure~\ref{fig:ring-w} (we will drop the index ``0'' for the sake of
brevity). The difference between the radial epicyclic frequency and the
orbital frequency gives the shift of pericentre. The difference of
vertical epicyclic and orbital frequencies gives the nodal precession.

Internal resonances can occur in the system
(\ref{eq:res_gov_r})--(\ref{eq:res_gov_theta}). In order to capture this
phenomenon, we carry out a multiple-scales expansion (Nayfeh \& Mook
\cite{nay79}) in the form
\begin{equation}
 \delta\rho(t,\epsilon)=\sum_{n=1}^4\epsilon^n\rho_n(T_\mu),\quad
 \delta \theta(t,\epsilon)=\sum_{n=1}^4\epsilon^n\theta_n(T_\mu),
\label{eq:ms-solexp}
\end{equation}
where $T_\mu=\epsilon^\mu t$ are treated as independent time scales. We
will terminate the expansion at the fourth order ($\mu=0,1,2,3$;
the number of time scales is the same as the order at which the
expansions are truncated). 

Time derivatives take a form of expansions
\begin{equation}
  \frac{\dd}{\dd t}=\sum_{\mu=0}^4\epsilon^\mu\D{\mu},
  \quad
  \frac{\dd^2}{\dd t^2}=\sum_{\mu=0}^4\sum_{\nu=0}^4\epsilon^{\mu+\nu}\D{\mu}\D{\nu},
\end{equation}
where $\D{\mu}\equiv\partial/\partial T_\mu$. The method tackles the governing 
equations in their general form. 

As a specific example we can adopt an explicit form describing the
orbital motion,
\begin{eqnarray}
 \delta\ddot{\rho}+\omega_r^2\,\delta\rho\!\!&=&\!\!
 (1+\delta\rho)\,\delta\dot{\theta}^2-
 \left[\frac{1}{r}\pder{\mathcal{U}}{r}-
 \omega_r^2\,\delta\rho\right],
 \label{eq:epc-nlinrh}\\
\delta\ddot{\theta}+\omega_\theta^2\,\delta\theta\!\!&=&\!\!
 -2\,\frac{\delta\dot{\rho}\,\delta\dot{\theta}}{1+\delta\rho}-
 \left[\frac{1}{(1+\delta\rho)^2}\pder{\mathcal{U}}{\theta}-
 \omega_\theta^2\,\delta\theta\right].
\label{eq:epc-nlinth}
\end{eqnarray}
Because equations (\ref{eq:epc-nlinrh})--(\ref{eq:epc-nlinth}) are
conservative, the growth of energy in one mode must be balanced by the
energy loss in the other mode. Close to resonance radii,
where the two epicyclic frequencies are in ratio of small integers, the
periodic exchange of energy should occur in a more pronounced rate.
Because amplitudes of the oscillations are connected with 
eccentricity and inclination, the solution alternates
between an inclined, almost circular trajectory at certain
stages, and an eccentric, almost equatorial case at some other time. We
remark that, in a non-linear system, the eigenfrequencies $\omega_r$, 
$\omega_\theta$ are expected to differ from the observed (i.e.\
predicted) frequencies, which can be revealed e.g.\ by Fourier analysis 
of data time series. Relevance of this fact for QPOs was first
recognized by Abramowicz et al.\ (\cite{abr03}) and Rebusco
(\cite{reb04}), when they discussed a model for Sco X-1. It was further
employed by Hor\'ak et al.\ (\cite{hor04b}), who suggested that a
connection should exist between the high-frequency QPOs and
normal-branch oscillations.

We expand the effective potential derivatives and the functions 
$f_\rho$ and $f_\theta$ into Taylor series, up to the fourth order about
a circular orbit. The expansion provides many nonlinear terms containing
various derivatives
\begin{equation}
u_{ij}\equiv\left(\frac{\partial^{i+j}\mathcal{U}}{\partial r^i\,\partial\theta^j}\right)_{[r_0,\pi/2]}.
\end{equation}
By imposing constraints on the potential and its derivatives, we
identify resonances that are expected in a particular system (and we reject
those resonances that cannot be realized). In the
next subsection, formulation of these constraints is still kept
completely general, valid for the arbitrary form of $\mathcal{U}$. Only
later, in subsection~\ref{subsection:constraints}, we come to our
original motivation from the orbital motion around a black hole and we
employ the symmetry of the potential $\mathcal{U}$. We consider
the case of potential symmetric with respect to the equatorial plane.
This implies the condition $u_{i(2k+1)}=0$, $k\in\mathcal{N}$, somewhat
reducing the number of terms in the expansions.

Amplitudes of the oscillations are characterized by a small parameter:
$\delta\rho\sim\epsilon$, $\delta\theta\sim\epsilon$. We
impose the solvability constraints and seek a solution in the form
(\ref{eq:ms-solexp}). To this aim, we first write the explicit form of
these constraints in different orders of approximation.

\subsection{Solvability constraints}
In the first order, we obtain equations describing two independent
harmonic oscillators,
\begin{equation}
 \left(\D{0}^2+\omega_r^2\right)\rho_1=0, \quad 
 \left(\D{0}^2+\omega_\theta^2\right)\theta_1=0.
 \label{eq:res_1}
\end{equation}
The solutions can be expressed in the form
\begin{equation}
 \rho_1=\A_\rho+\A_{-\rho}, \quad \theta_1=\A_\theta+\A_{-\theta},
 \label{eq:res_1sol}
\end{equation}
where we denote $\A_x\equiv A_x e^{\ii \omega_x T_0}$, $\A_{-x}=A_x^\ast
e^{-\ii \omega_x T_0}$ with $x=\rho$ or $\theta$. The complex amplitudes
$A_x$ generally depend on slower scales, $T_1$, $T_2$ and $T_3$. 

An algorithmic nature of the multiple-scales method allows us
to determine the form of solvability conditions in a conservative system
with two degrees of freedom. The conditions arise by eliminating the
terms that are secular in the fastest variable, $T_0$. This constraint
is imposed, because otherwise the solutions in the form of series would
not converge uniformly. In fact, the reason why the same number of 
scales is required as the order of approximation in the 
expansion~(\ref{eq:ms-solexp}) is that one secular term gets eliminated 
in each order, and therefore the functions $A_x(T_\mu)$ are determined by 
the same number of solvability conditions as the number of their variables.

\begin{table}[t]
\caption{Possible resonances and secular terms in the second order of
approximation. Only regular secular terms are present if $\omega_\theta$
and $\omega_r$ are outside resonance, as indicated in the left column.
Subsequent lines refer to a system in $1$:$2$ and $2$:$1$ resonances,
respectively. In each record corresponding to the resonances, the
first/second row gives an expression for secular terms in
radial/vertical oscillations. To simplify our notation we introduced
$\Lambda_\alpha\equiv C^{(\rho)}_\alpha$ and $\Kappa_\alpha\equiv
C^{(\theta)}_\alpha$.}
\label{tab:res_2}
\begin{center}
\begin{tabular}{c@{~~~}l}
  \hline\hline
  \rule{0mm}{3.5ex} \rule[-3ex]{0mm}{3.5ex}
  $\omega_\theta $:$ \omega_r$  &
    ~~~~~~~{Secular terms}
  \\
  \hline
  \rule{0mm}{3.5ex}
  \textrm{Outside} &
    $-2\ii\omega_r\D{1}\A_\rho$
  \\
  \textrm{resonance} &
    $-2\ii\omega_\theta\D{1}\A_\theta$
  \rule[-3ex]{0mm}{3.5ex}
  \\
  \hline
  \rule{0mm}{3.5ex}
  $1$:$2$ &
    $-2\ii\omega_r\D{1}\A_\rho$,~~$\Kappa_{1001}\A_{-\rho}\A_\theta$
  \\
  ~ &
    $-2\ii\omega_\theta\D{1}\A_\theta$,~~$\Lambda_{0200}\A_\rho^2$
  \rule[-3ex]{0mm}{3.5ex}
  \\
  \hline
  \rule{0mm}{3.5ex}
  $2$:$1$ &
    $-2\ii\omega_r\D{1}\A_\rho$,~~$\Kappa_{0002}\A_\theta^2$
  \\
  ~ &
    $-2\ii\omega_\theta\D{1}\A_\theta$,~~$\Lambda_{0110}\A_\rho\A_{-\theta}$
  \rule[-3ex]{0mm}{3.5ex}
  \\
  \hline
\end{tabular}
\end{center}
\end{table}

In the second order, the terms proportional to $\epsilon^2$ in the
expanded left-hand side of the governing equations 
(\ref{eq:res_gov_r})--(\ref{eq:res_gov_theta}) are
\begin{equation}
 \left[\ddot{\delta x}+\omega_x^2 x\right]_2=
 \left(\D{0}^2+\omega_x^2\right)
 x_2+2\ii\omega_x\D{1}\A_x-2\ii\omega_x\D{1}\A_{-x}.
\label{eq:res_2lhs}
\end{equation}
On the right-hand side, the second-order terms result from the expansion
of the nonlinearity
$f_x(\delta\rho,\delta\theta,\dot{\delta\rho},\dot{\delta\theta})$ with
$\D{0}\rho_1$ and $\D{0}\theta_1$ in place of $\dot{\delta\rho}$ and
$\dot{\delta\theta}$, respectively. They can be expressed as linear
combinations of quadratic terms constructed from $\A_{\pm\rho}$ and
$\A_{\pm \theta}$:
\begin{equation}
 \left[ f_x(\delta\rho, \delta\theta, \dot{\delta\rho}, \dot{\delta \theta})
 \right]_2 = \sum_{|\alpha|=2} C_\alpha^{(x)}
 \A_{-\rho}^{\alpha_1}\A_{\rho}^{\alpha_2}\A_{-\theta}^{\alpha_3}\A_{\theta}^{\alpha_4},
\label{eq:res_2rhs}
\end{equation}
where $\alpha=(\alpha_1,\ldots,\alpha_4)$ and
$|\alpha|=\alpha_1+\ldots+\alpha_4$. The constants $C_\alpha^{(x)}$ are
given by angular frequencies of $\omega_x$ and by coefficients of the
Taylor expansion of $f_x$. Equating the right-hand sides of eqs.\
(\ref{eq:res_2lhs})--(\ref{eq:res_2rhs}) we find
\begin{eqnarray}
 \left(\D{0}^2 + \omega_x^2\right) x_2 &=& - 2\ii\omega_x \D{1}\A_x  +
 2\ii\omega_x\D{1}\A_{-x} 
\nonumber\\
&&+\sum_{|\alpha|=2} C_\alpha^{(x)}
 \A_{-\rho}^{\alpha_1}\A_{\rho}^{\alpha_2}
 \A_{-\theta}^{\alpha_3}\A_{\theta}^{\alpha_4}.
\label{eq:res_2}
\end{eqnarray} 
The rhs of eq.~(\ref{eq:res_2}) contains one secular term independently
of the eigenfrequencies $\omega_r$ and $\omega_\theta$. However,
additional secular terms may appear in the resonance. For example, when
$\omega_r \approx 2 \omega_\theta$, the terms proportional to
$\A^2_\theta$ in the $\rho$-equation ($x\rightarrow\rho$) and 
$\A_\rho\A_{-\theta}$ in the $\theta$-equation ($x\rightarrow\theta$)
become secular and they should be also included to the solvability
conditions. The similar situation happens when
$\omega_r\approx\omega_\theta/2$. These are internal resonances, which
show a qualitatively different behaviour: the corresponding terms are
secular only for special (resonant) combinations of $\omega_r$ and
$\omega_\theta$, contrary to the terms that appear always and are
referred to as regular secular terms. Possible resonances in the second
order of approximation and the corresponding secular terms in
eq.~(\ref{eq:res_2}) are listed in Table~\ref{tab:res_2}. Let us assume,
for a moment, that the system is far from any resonance. Then
\begin{equation}
 \label{eq:res_2sec}
\D{1} A_x = 0.
\end{equation}
The frequencies and amplitudes are constant and the behaviour of the
system is almost identical as what one finds in the linear
approximation. The only difference is the presence of higher harmonics
oscillating with frequencies $2\omega_r$, $2\omega_\theta$ and 
$|\omega_r\pm\omega_\theta|$. They are given by a particular solution of
equation (\ref{eq:res_2}) after eliminating the secular term,
\begin{equation}
 \label{eq:res_2sol}
 x_2 = \sum_{|\alpha|=2} Q_\alpha^{(x)}
 \A_{-\rho}^{\alpha_1}\A_{\rho}^{\alpha_2}
 \A_{-\theta}^{\alpha_3}\A_{\theta}^{\alpha_4},
\end{equation}
Under the assumption of time-reflection invariance, constants
$Q_\alpha^{(x)}$ are real and their relation to constants
$C_\alpha^{(x)}$ becomes obvious by substituting $x_2$ into equation
(\ref{eq:res_2}). We find
\begin{equation}
 Q^{(x)}_{klmn}=\frac{C^{(x)}_{klmn}}{\omega_x^2-\left[(k-l)\omega_r+
 (m-n)\omega_\theta\right]^2}\,.
\end{equation}

\begin{table}[t]
\caption{Possible resonances and secular terms in the third order of 
approximation. Individual records have a similar meaning as in Tab.~1.}
\label{tab:res_3}
\begin{center}
\begin{tabular}{c@{~~~}l}
  \hline\hline\rule{0ex}{3ex}\rule[-2.5ex]{0ex}{3ex}
  $\omega_\theta$:$\omega_r$  &
    ~~~~~~~{Secular terms}
  \rule{0mm}{3.5ex}
  \\
  \hline
  \rule{0mm}{3.5ex}
  \textrm{Outside}&
    $2\ii\omega_r\D{2}\A_\rho$,~~$\Kappa_{1200}|A_\rho|^2\A_\rho$,~~$\Kappa_{0111}|A_\theta^2|\A_\rho$
  \\
  \textrm{resonance}&
    $2\ii\omega_\theta\D{2}\A_\theta$,~~$\Lambda_{1101}|A_\rho|^2\A_\theta$,~~$\Lambda_{0012}|A_\theta^2|\A_\theta$
  \rule{0mm}{3.5ex}
  \\
  \hline
  \rule{0mm}{3.5ex}
  $1$:$3$ &
    $2\ii\omega_r\D{2}\A_\rho$,~~$\Kappa_{1200}|A_\rho|^2\A_\rho$,~~$\Kappa_{0111}|A_\theta^2|\A_\rho$,~~$\Kappa_{0030}\A_\theta^3$
  \\
  ~&
    $2\ii\omega_\theta\D{2}\A_\theta$,~~$\Lambda_{1101}|A_\rho|^2\A_\theta$,~~$\Lambda_{0012}|A_\theta^2|\A_\theta$,~~$\Lambda_{0120}
\A_\rho\A_{-\theta}^2$
  \rule{0mm}{3.5ex}
  \\
  \hline
  \rule{0mm}{3.5ex}
  $1$:$1$ &
    $2\ii\omega_r\D{2}\A_\rho$,~~$\Kappa_{1200}|A_\rho|^2\A_\rho$,~~$\Kappa_{0111}|A_\theta^2|\A_\rho$,
    ~~$\Kappa_{1110}|A_\rho|^2\A_\theta$,
  \\
  ~&
    $\Kappa_{0012}|A_\theta|^2\A_\theta$,~~$\Kappa_{0210}\A_\rho^2\A_{-\theta}$,~~$\Kappa_{1002}\A_{-\rho}\A_\theta^2$
  \\
  \rule{0ex}{2.5ex}
  ~&
    $2\ii\omega_\theta\D{2}\A_\theta$,~~$\Lambda_{1101}|A_\rho|^2\A_\theta$,~~$\Lambda_{0012}|A_\theta^2|\A_\theta$,
    ~~$\Lambda_{2100}|A_\rho|^2\A_\rho$,
  \\
  ~&
    $\Lambda_{0021}|A_\theta|^2\A_\theta$,~~$\Lambda{1002}\A_{-\rho}\A_\theta^2$,~~$\Lambda_{0210}\A_\rho^2\A_{-\theta}$
  \rule{0mm}{3.5ex}
  \\
  \hline\rule{0ex}{3ex}
  $ 3$:$1 $ &
    $2\ii\omega_r\D{2}\A_\rho$,~~$\Kappa_{1200}|A_\rho|^2\A_\rho$,~~$\Kappa_{0111}|A_\theta^2|\A_\rho$,
    ~~$\Kappa_{2001}\A_{-\rho}^2\A_\theta$
  \\
  ~&
    $2\ii\omega_\theta\D{2}\A_\theta$,~~$\Lambda_{1101}|A_\rho|^2\A_\theta$,~~$\Lambda_{0012}|A_\theta^2|\A_\theta$,
    ~~$\Lambda_{0300}\A_\rho^3$
  \rule{0mm}{3ex}
  \\
  \hline
\end{tabular}
\end{center}
\end{table}

In the third order, the discussion is analogous in many respects. The
terms proportional to $\epsilon^3$, which appear on the lhs of the
governing equations, are given by
\begin{equation}
 \left[ \ddot{\delta x} + \omega_x^2 x \right]_3 =
 \left(\D{0}^2 + \omega_x^2\right) x_3  +  2\ii\omega_x\D{2}\A_x  -
  2\ii\omega_x\D{2}\A_{-x}.
\end{equation}
The terms containing $D_1 x_1$ and $D_1 x_2$ vanish in consequence of
the solvability condition (\ref{eq:res_2sec}). The rhs contains now
cubic terms of the Taylor expansion. We obtain
\begin{eqnarray}
 \left(\D{0}^2 + \omega_x^2\right)x_3&=&-2\ii\omega_x \D{2}\A_x +
 2\ii\omega_x\D{2}\A_{-x}
\nonumber\\
&&+\sum_{|\alpha|=3} C_\alpha^{(x)}
 \A_{-\rho}^{\alpha_1}\A_{\rho}^{\alpha_2}
 \A_{-\theta}^{\alpha_3}\A_{\theta}^{\alpha_4},
\label{eq:res_3}
\end{eqnarray}
where constants $C_\alpha^{(x)}$ are real. The secular terms are
summarized in Table~\ref{tab:res_3} together with resonances possible in
the third order of approximation. Again, far from any resonance we
eliminate the terms that are secular independently of $\omega_r$,
$\omega_\theta$. The resulting solvability conditions take the form
\begin{eqnarray}
 \D{2} A_\rho = -\frac{\ii}{2\omega_r}\left[\Kappa_{1200}|A_\rho|^2 +
 \Kappa_{0111}|A_\theta|^2 \right]A_\rho,
 \label{eq:res_3sec_r}\\
 \D{2} A_\theta = -\frac{\ii}{2\omega_\theta}\left[\Lambda_{1101}|A_\rho|^2 +
 \Lambda_{0012}|A_\theta|^2 \right]A_\theta.
\label{eq:res_3sec_m}
\end{eqnarray}
A particular solution of eq.~(\ref{eq:res_3}) is given by a linear
combination of cubic terms constructed from $\A_{\pm\rho} $ and 
$\A_{\pm\theta}$,
\begin{equation}
 x_3 = \sum_{|\alpha|=3} Q_\alpha^{(3,x)}
 \A_{-\rho}^{\alpha_1}\A_{\rho}^{\alpha_2}
 \A_{-\theta}^{\alpha_3}\A_{\theta}^{\alpha_4},
 \label{eq:res_3sol}
\end{equation}
where all coefficients $Q_\alpha^{(3,x)}$ are now real.

\begin{table}[t]
\caption{Possible resonances in the fourth order of approximation.}
\label{tab:res_4}
\begin{center}
\begin{tabular}{c@{~~~}l}
  \hline\hline
  \rule{0ex}{3ex}\rule[-2.5ex]{0ex}{3ex}
  $ \omega_\theta $:$ \omega_r $  &
    ~~~~~~~{Secular terms}
  \\
  \hline
  \textrm{Outside}&\rule{0ex}{3ex}
    $2\ii\omega_r\D{3}\A_\rho$
  \\
  \textrm{resonance}&
    $2\ii\omega_\theta\D{3}\A_\theta$
  \\
  \hline
  \rule{0ex}{3ex}
  $1$:$4$ &
    $2\ii\omega_r\D{3}\A_\rho$,~~$\Kappa_{0004}\A_\theta^4$
  \\
  ~&
    $2\ii\omega_\theta\D{3}\A_\theta$,~~$\Lambda_{0103}\A_\rho\A_\theta^3 $
  \rule[-2.5ex]{0ex}{3ex}
  \\
  \hline
  \rule{0ex}{3ex}
  $2$:$3$ &
    $2\ii\omega_r\D{3}\A_\rho$,~~$\Kappa_{0130}\A_\rho\A_{-\theta}^3$
  \\
  ~&
    $2\ii\omega_\theta\D{3}\A_\theta$,~~$\Lambda_{0220}\A_\rho^2\A_{-\theta}^2$
    \rule[-2.5ex]{0ex}{3ex}
  \\
  \hline
  \rule{0ex}{3ex}
  $3$:$2$ &
    $2\ii\omega_r\D{3}\A_\rho$,~~$\Kappa_{2002}\A_{-\rho}^2 \A_\theta^2$
  \\
  ~&
    $2\ii\omega_\theta\D{3}\A_\theta$,~~$\Lambda_{0310}\A_\rho^3\A_{-\theta} $
  \rule[-2.5ex]{0ex}{3ex}
  \\
  \hline
  \rule{0ex}{3ex}
  $4$:$1$ &
    $2\ii\omega_r\D{3}\A_\rho$,~~$\Kappa_{0301} \A_\rho^3\A_\theta$
  \\
  ~&
    $2\ii\omega_\theta\D{3}\A_\theta$,~~$\Lambda_{0400}A_\rho^4 $
  \rule[-2.5ex]{0ex}{3ex}
  \\
  \hline
\end{tabular}
\end{center}
\end{table}

Finally, in the fourth order of the approximation,
\begin{equation}
 \left[\ddot{\delta x}+\omega_x^2 x \right]_4=
 \left(\D{0}^2+\omega_x^2\right)
 x_3+2 \D{3}\D{0}x_1+2\D{0}\D{2}x_2.
\end{equation}
The operator $\D{0}\D{2}$ acts on $x_2$, given by
eq.~(\ref{eq:res_2sol}). The resulting form is found by employing the
solvability conditions (\ref{eq:res_3sec_r}) and (\ref{eq:res_3sec_m}):
\begin{equation}
 2 \D{0} \D{2} x_2 = \omega_x^2 \sum_{|\alpha| = 4} J_\alpha^{(x)}
 \A_{-\rho}^{\alpha_1}\A_{\rho}^{\alpha_2}
 \A_{-\theta}^{\alpha_3}\A_{\theta}^{\alpha_4},
\label{eq:res_3lhs}
\end{equation}
where $J^{(x)}_\alpha$ are real constants. By expanding the rhs we
arrive at the governing equation
\begin{eqnarray}
 \left[\D{0}^2 + \omega_x^2\right]x_4&=&-2\ii\omega_x \D{3}\A_x+
 2\ii\omega_x\D{3}\A_{-x}
\nonumber\\
&&+\sum_{|\alpha|=4} C_\alpha^{(x)}
 \A_{-\rho}^{\alpha_1}\A_{\rho}^{\alpha_2}
 \A_{-\theta}^{\alpha_3}\A_{\theta}^{\alpha_4},
\label{eq:res_4}
\end{eqnarray}
with $C_\alpha^{(x)}$ real constants. Only one secular term independent
of $\omega_r$ and $\omega_\theta$ appears on the rhs: $-2\ii\omega_x
\D{3}\A_x$. The sum contains only terms that become secular near a
resonance. These terms and the solvability conditions are listed in
Table~\ref{tab:res_4}.

A notable feature of internal resonances $k$:$l$ is that $k\omega_r$
and $l\omega_\theta$ need not be infinitesimally close to each other,
as might be expected from the linear analysis. Consider, for example, an
internal resonance $1$:$2$, i.e.\ $\omega_\theta\approx2\omega_r$. By
eliminating the secular terms, we obtain solvability conditions (see
Tab.~\ref{tab:res_2})
\begin{eqnarray}
-2\ii\omega_r\D{1}\A_\rho + \Kappa_{1001}\A_{-\rho}\A_\theta&=&0,\\
-2\ii\omega_\theta\D{1}\A_\theta +  \Lambda_{0200}\A_\rho^2&=&0.
\end{eqnarray}
In each of these equations the first term is regular, while the second
term is nearly secular (resonant) one. The solvability conditions give
us the long-term behaviour of the amplitudes and phases of oscillations.
Suppose now that the system departs from the sharp ratio by small
(first-order) deviations $\omega_\theta=2\omega_r+\epsilon\sigma$, where
$\sigma$ is the detuning parameter. The terms
proportional to $\A_{-\rho}\A_\theta$ and $\A_\rho^2$ still remain
secular with respect to the variable $T_0$. This can be demonstrated
from
$\A_{-\rho}\A_\theta = A_\rho^\ast A_\theta e^{\ii (\omega_\theta - \omega_r)
 T_0} = A_\rho^\ast A_\theta e^{\ii \sigma T_1} e^{\ii \omega_r T_0}$.
Similar relation holds for $\A_\rho^2$.

\subsection{Solution of the solvability constraints}
\label{subsection:constraints}
By comparing the coefficients with same powers of $\epsilon$ on both sides
of the Taylor-expanded governing equations 
(\ref{eq:ms-generalfrh})--(\ref{eq:ms-generalfth}), we obtain relations
for functions $\rho_i(T_j)$ and $\theta_i(T_j)$ that can be solved
successively. After rearranging to a ``canonical'' form,
\begin{eqnarray}
 \left[\D{0}^2+\omega_r^2\right]\rho_n&=&\sum\,\Kappa_{ijkl}
 \A_{-\rho}^i\A_{\rho}^j\,\A_{-\theta}^k\A_{\theta}^l,
 \label{eq:epc-formrh}\\
 \left[\D{0}^2+\omega_\theta^2\right]\theta_n&=&\sum\,\Lambda_{ijkl}
 \A_{-\rho}^i\A_{\rho}^j\,\A_{-\theta}^k\A_{\theta}^l,
\label{eq:epc-formth}
\end{eqnarray}
where $n$ is the order of approximation. In this way we identify
constants $K_{ijkl}$ and $\Lambda_{ijkl}$. The studied gravitational
potential is symmetric with respect to the equatorial plane, therefore,
the series (\ref{eq:epc-formrh})--(\ref{eq:epc-formth}) cannot contain
terms proportional to odd derivatives of the effective potential with
respect to $\theta$. Hence, contrary to a general case, only  specific
resonances occur here: $\omega_\theta$:$\omega_r=1$:$2$, $1$:$1$,
$3$:$2$ and $1$:$4$. These are all possible combinations that may
occur within the given order of approximation (the first three cases
were originally identified by Rebusco \cite{reb04}, although she does
not mention the fourth possible combination). A general argument
of non-linear analysis suggests that the dominant resonances are those
ones which correspond to ratios of small natural numbers, although not
every conceivable resonant combination comes up in a given physical
system. Indeed, it appears that the $3$:$2$ ratio is the most important case
when the high-frequency QPO pairs are debated, however, the true role of this
resonance has not yet been understood; see also the discussion in Bursa
(\cite{bur05}) and Lasota (\cite{las05}). Proceeding further to higher-order 
terms of the expansions reveals even more resonances, but
these are expected to be very weak (recently various kinds of weird
combinations have been examined by T\"or\"ok et al.\ \cite{tor05}).
Hereafter we concentrate on the first three combinations.

\subsubsection*{The case of 1:2 resonance} 
The solvability conditions take the form (cp.\ Tab.~\ref{tab:res_2} and
Hor\'ak \cite{hor04})
\begin{eqnarray}
 \D{1}\A_\rho &=&-\frac{\ii}{2\omega_r}\Kappa_{0002}\A_\theta^2,
 \label{eq:nepc-solv12rh}\\
 \D{1}\A_\theta &=&-\frac{\ii}{2\omega_\theta}\Lambda_{0110}\A_\rho\,\A_{-\theta},
\label{eq:nepc-solv12th}.
\end{eqnarray}
The coefficients of the resonant terms are given by
\begin{equation}
 \Kappa_{0002}=-\omega_\theta^2-\frac{u_{12}}{2r_0},
\quad
 \Lambda_{0110}=-2\omega_\theta^2-\frac{u_{12}}{r_0},
\label{eq:nepc-coeff12}
\end{equation}
satisfying a mutual relation $2\Kappa_{0002}=\Lambda_{0110}$.

\subsubsection*{The case of 1:1 resonance}
The solvability conditions for the first order,
$\D{1}\A_\rho=\D{1}\A_\theta=0$, imply that the complex amplitudes
$\A_\rho$ and $\A_\theta$ depend only on the second scale $T_2$. The
$1$:$1$ ($\omega_r\approx\omega_\theta$) resonance is the only epicyclic
resonance of the system with reflection symmetry, which occurs in the
third order of approximation. The dependence on $T_2$ implies a slower
behaviour. The solvability conditions are
\begin{eqnarray}
 \D{2}\A_\rho&=&-\frac{\ii}{2\omega_r}\Big[
 \Kappa_{1200}\left|A_\rho\right|^2\A_\rho 
\nonumber\\
&&+\Kappa_{0111}\left|A_\theta\right|^2\A_\rho 
 +\Kappa_{1002}\A_{-\rho}\,\A_\theta^2 \Big],
\label{eq:nepc-solv11rh}\\
 \D{2}\A_\theta&=&-\frac{\ii}{2\omega_\theta}\Big[
 \Lambda_{1101}\left|A_\rho\right|^2\A_\theta 
\nonumber\\
&&+\Lambda_{0012}\left|A_\theta\right|^2\A_\theta 
 + \Lambda_{0210}\A_\rho^2\A_{-\theta} \Big]
\label{eq:nepc-solv11th}
\end{eqnarray}
and the coefficients of the resonant terms are given by
\begin{eqnarray}
 \Kappa_{1200}&=& r_0^2\left(\frac{5\,u_{30}^2}{6\,\omega_\theta}-
 \ratio{1}{2}u_{40}\right),
\label{eq:nepc-11K1200}\\
 \Kappa_{0111}&=& \frac{1}{3}\left(-10\,\omega_\theta^2 +
 \frac{2u_{12}^2}{r_0^2\omega_\theta^2} - 3\,u_{22} \right.
\nonumber\\
&& \left. - 6\,r_0u_{30} +
 u_{12}\left[\frac{8}{r_0}+\frac{3\,u_{30}}{\omega_\theta^2}\,\right]\right),
 \\
 \Kappa_{1002}&=& \frac{1}{6}\left(-6\,\omega_\theta^2 +
 \frac{6\,u_{12}^2}{r_0^2\omega_\theta^2} - 3\,u_{22} - 2\,r_0u_{30} -
 \frac{u_{12}\,u_{30}}{\omega_\theta^2}\right),
 \\
 \Lambda_{0012}&=&-\frac{u_{04}}{2r_0^2}-
 \frac{7\,u_{12}}{6\,r_0}+\frac{5\,u_{12}^2}{6\,r_0^2\omega_\theta^2}+
 \frac{10}{3}\omega_\theta^2,
 \\
 \Lambda_{0210}&=&\Kappa_{1002},
 \\
 \Lambda_{1101}&=&\Kappa_{0111}.
\label{eq:nepc-11L0111}
\end{eqnarray}
					
\subsubsection*{The case of 3:2 resonance}
The solvability conditions involve both the third and the fourth orders.
Hence, the amplitudes $A_\rho$, $A_\theta$ are functions of both time
scales $T_3$ and $T_4$. The elimination of regular secular terms in the
third order ($3\omega_r\approx2\omega_\theta$; see Tab.~\ref{tab:res_3})
gives
\begin{eqnarray}
 \D{2}\A_\rho&=&-\frac{\ii}{2\omega_r}\left[
 \Kappa_{1200}\left|A_\rho\right|^2\A_\rho +
 \Kappa_{0111}\left|A_\theta\right|^2\A_\rho\right],
 \\
 \D{2}\A_\theta&=&-\frac{\ii}{2\omega_\theta}\left[
 \Lambda_{1101}\left|A_\rho\right|^2\A_\theta +
 \Lambda_{0012}\left|A_\theta\right|^2\A_\theta\right].
\end{eqnarray}
with the coefficients
\begin{eqnarray}
 \Kappa_{1200}&=& r_0^2\left(\frac{15\,u_{30}^2}{8\,\omega_\theta}-
 \ratio{1}{2}u_{40}\right),
 \\
 \Kappa_{0111}&=&\frac{1}{4}\left(-15\,\omega_\theta^2 +
 \frac{9\,u_{12}}{4\,r_0^2\omega_\theta^2}-4\,u_{22} \right.
\nonumber\\
&& \left. - 18\,r_0 u_{30} +
 9\,u_{12}\left[\frac{1}{r_0}+\frac{u_{30}}{\omega_\theta^2}\right]\right),
 \\
 \Lambda_{0012}&=&-\frac{u_{04}}{2r_0^2} +
 \frac{135}{64}\,\frac{u_{12}^2}{r_0^2\omega_\theta^2} -
 \frac{153}{16}\,\frac{u_{12}}{r_0} + \frac{135}{16}\,\omega_\theta^2,
 \\
 \Lambda_{1101}&=&\Kappa_{0111}.
\end{eqnarray}
The elimination of the resonant terms gives the solvability condition in
the fourth order (Tab.~\ref{tab:res_4}),
\begin{equation}
\D{3}\A_\rho=-\frac{\ii}{2\omega_r}\,\Kappa_{2002}\A_{-\rho}^2\A_{\theta}^2,
\quad
\D{3}\A_\theta=-\frac{\ii}{2\omega_\theta}\,\Lambda_{0310}\A_\rho^3\A_{-\theta},
\end{equation}
where the resonant coefficients are
\begin{eqnarray}
 &&\Kappa_{2002}=-\ratio{15}{16}\,\omega_\theta^2 + \ratio{27}{32}\,\frac{u_{12}}{r_0}+
 \ratio{135}{64}\,\frac{u_{12}^2}{r_0^2\omega_\theta^2} - \ratio{243}{128}\,
 \frac{u_{12}^3}{r_0^3\omega_\theta^4}
 \nonumber\\
 &&-\ratio{9}{8}\,u_{22} + \ratio{27}{16}\,
 \frac{u_{12}\,u_{22}}{r_0\omega_\theta^2}
 -\ratio{27}{16}\,r_0u_{30} + \ratio{81}{64}
 \frac{u_{12}^2\,u_{30}}{r_0\omega_\theta^4}
 \nonumber\\
 &&
 -\ratio{9}{16}\,\frac{r_0u_{22}\,u_{30}}{\omega_\theta^2}-\ratio{81}{256}\,
 \frac{r_0^2u_{30^2}}{\omega_\theta^2}
 -\ratio{81}{512}\,\frac{r_0u_{12}\,u_{30}^2}{\omega_\theta^4}
 \nonumber\\
 &&
 -\ratio{1}{4}\,r_0u_{30}-
 \ratio{9}{64}\,r_0^2u_{40}-\ratio{9}{128}\,\frac{r_0u_{12}\,u_{40}}{\omega_\theta^2},
 \\
 &&\Lambda_{0310}=\ratio{2}{3}\Kappa_{2002}.
\end{eqnarray}
By introducing the detuning parameter,
\begin{equation}
 \sigma\equiv3\,\frac{\omega_r}{\omega_\theta}-2=\epsilon^2\tilde{\sigma}_2+\epsilon^3\tilde{\sigma}_3,
\end{equation}
the solvability conditions adopt the explicit form
\begin{eqnarray}  
 2\ii\omega_r\D{3}A_\rho\!&=&\!
 \Kappa_{2002}\left(A_\rho^\star\right)^2 A_\theta^2\,\,e^{\ii(\tilde{\sigma}_2T_2+\tilde{\sigma}_3T_3)},
 \\
 2\ii\omega_\theta\D{3} A_\theta\!&=&\!
 \Lambda_{0310}A_\rho^3 A_\theta^\star\,\,e^{\ii\left(\tilde{\sigma}_2T_2+\tilde{\sigma}_3T_3\right)},
 \\
 2\ii\omega_r\D{2}A_\rho\!&=&\!
 \left[\Kappa_{1200}\left|A_\rho\right|^2 + \Kappa_{0111}\left|A_\theta\right|^2\right]A_\rho,
 \\
 2\ii\omega_\theta\D{2} A_\theta\!&=&\!
 \left[\Lambda_{1101}\left|A_\rho\right|^2 + \Lambda_{0012}\left|A_\theta\right|^2\right]A_\theta.
\end{eqnarray}

\begin{figure}[t]
\begin{center}
\includegraphics[width=0.5\textwidth]{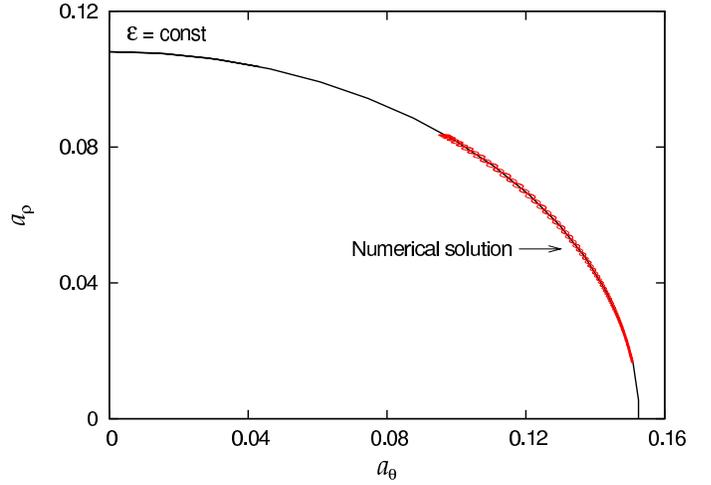}
\end{center}
\caption{Comparison 
between the analytical constraint $\mathcal{E}(a_\rho,a_\theta)=\const$ 
(an ellipse), derived in multiple-scales approximation, and the
corresponding exact (numerical) solution. 
The curly curve is the numerical solution of the oscillation
amplitudes $a_\rho(t)$, $a_\theta(t)$. The agreement between the two
curves demonstrates that accuracy of the approximation is satisfactory
over the entire time span. See the text for details.}
\label{fig:qpo-ellipse}
\end{figure}
\noindent
Finally, by substituting a polar form of complex amplitudes, 
$A_\rho\equiv \frac{1}{2}\tilde{a}_\rho e^{\ii\phi_\rho}$ and
$A_\theta\equiv \frac{1}{2}\tilde{a}_\theta e^{\ii\phi_\theta}$, we
get a set of eight equations governing the long-term behaviour of 
phases and amplitudes:
\begin{eqnarray}
 \D{2}\tilde{a}_\rho\!\!&=&\!\!0,
 \quad
 \D{2}\tilde{a}_\theta=0,
\label{eq:D2a}\\
 \D{3}\tilde{a}_\rho\!\!&=&\!\!
 \frac{\Kappa_{2002}}{16\omega_r}\tilde{a}_\rho^2\tilde{a}_\theta^2 \sin\gamma,
 \quad
 \D{3}\tilde{a}_\theta=
 -\frac{\Lambda_{0310}}{16\omega_\theta}\tilde{a}_\rho^3\tilde{a}_\theta \sin\gamma,
\label{eq:D3a}\\
 \D{2}\phi_\rho\!\!&=&\!\!
 -\frac{1}{8\omega_r}\left[\Kappa_{1200}\tilde{a}_\rho^2 + \Kappa_{0111}\tilde{a}_\theta^2\right],
 \nonumber\\
 \D{2}\phi_\theta\!\!&=&\!\!
 -\frac{1}{8\omega_\theta}\left[\Lambda_{1101}\tilde{a}_\rho^2 + \Lambda_{0012}\tilde{a}_\theta^2\right],
\label{eq:D2phi}\\
 \D{3}\phi_\rho\!\!&=&\!\!
 -\frac{\Kappa_{2002}}{16\omega_r}\,\tilde{a}_\rho\tilde{a}_\theta^2\cos\gamma,
 \quad
 \D{3}\phi_\theta=
 -\frac{\Lambda_{0310}}{16\omega_\theta}\,\tilde{a}_\rho^3\cos\gamma,
\label{eq:D3phi}
\end{eqnarray}
where the phase function has been introduced as 
$\gamma(T_2,T_3)\equiv-\sigma_2T_2-\sigma_3T_3-3\phi_\rho+2\phi_\theta$.
The amplitudes $\tilde{a}_\rho$ and $\tilde{a}_\theta$ of the
oscillations vary slowly, because they depend only on the third
time-scale $T_3$. Phases $\phi_r$ and $\phi_\theta$ of the oscillations
evolve on both time scales $T_2$ and $T_3$.

\begin{figure}[t]
\begin{center}
\includegraphics[width=0.5\textwidth]{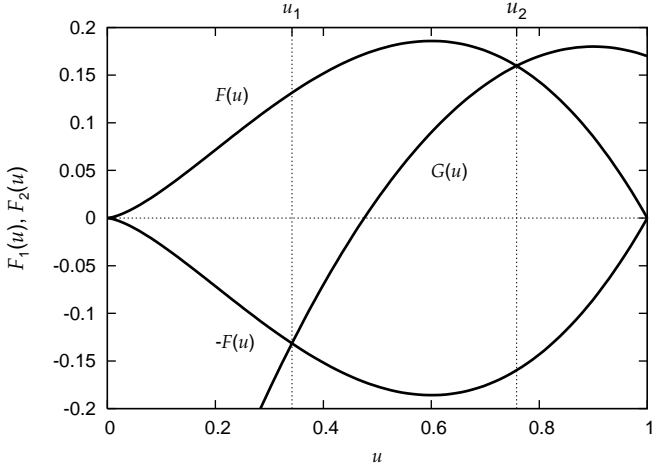}
\end{center}
\caption{The functions $F_1\,\equiv\,F(u)$, $F_2\,\equiv\,-F(u)$, and $G(u)$ from
eq.~(\ref{eq:nepc-32u}). The system evolution is limited to the interval
${\langle}u_1,u_2{\rangle}$, where the condition $|F(u)|\geq|G(u)|$ is
satisfied.}
\label{fig:nepc-fg32}
\end{figure}

\section{The system evolution near the 3:2 resonance}
\subsection{The integrals of motion}
The case of $3$:$2$ resonance is particularly relevant for the
high-frequency QPOs, both on observational and theoretical grounds (cf.\
Abramowicz \& Klu\'zniak \cite{abr01}; Klu\'zniak et al.\ \cite{klu04}
for arguments in favour of $3$:$2$  ratio in high-frequency QPOs and for
further references).  We therefore discuss this case in more detail,
however, similar discussion could be presented also for other resonances
(Hor\'ak \cite{hor05}). Reintroducing single physical time $t$, the
equations for the  second and the third order can be combined together.
Time derivatives are then given by
${\dd/\dd}t=\epsilon^2D_2+\epsilon^3D_3$. Amplitudes and  phases of the
oscillations are governed by equations
\begin{eqnarray}
 \dot{a}_r &=&
 \ratio{1}{24}\beta \omega_r a_r^2 a_\theta^2\sin\gamma,
 \label{eq:nepc-32arh}
 \\
 \dot{a}_\theta &=&
 -\ratio{1}{16}\beta\omega_\theta a_r^3 a_\theta\sin\gamma,
 \label{eq:nepc-32ath}
 \\
 \dot{\gamma} &=&
 -\sigma\omega_\theta + \frac{\omega_\theta}{4}\left[\mu_r a_r^2 +
 \mu_\theta a_\theta^2 + \frac{a_r}{2}\left( \alpha a_\theta^2 - \beta
 a_r^2 \right) \cos \gamma \right],
 \label{eq:nepc-32ga}
\end{eqnarray}
where $a_\rho=\epsilon\tilde{a}_\rho$, $a_\theta=\epsilon\tilde{a}_\theta$ 
and $\mu_r$, $\mu_\theta$ and $\beta$ are defined by relations
$\Lambda_{0310}=\ratio{2}{3}\Kappa_{2002}=\beta\omega_\theta^2$, $
\Kappa_{1200}-\Lambda_{1101}=\omega_r^2\mu_r$, and
$\Kappa_{0111}-\Lambda_{0012}=\omega_\theta^2\mu_\theta$. The
amplitudes and phases are not mutually independent; the equations
(\ref{eq:nepc-32arh})--(\ref{eq:nepc-32ath}) imply that the quantity
\begin{equation}
 \mathcal{E}=a_\rho^2 + \ratio{9}{4}a_\theta^2 = \const
 \label{eq:qpos-energy}
\end{equation}
remains conserved during the system evolution. Clearly, $\mathcal{E}$ is
proportional to the total energy of the oscillations. The existence of
this integral is a general property of a conservative system. Naturally,
it is not limited to the particular form of the  perturbing potential
(\ref{eq:epc-ringPhi}), which we consider here,  and it holds in
Newtonian, pseudo-Newtonian as well as  general-relativity versions of
the equations of motion (the  pseudo-Newtonian case was examined, in
detail, by Abramowicz et al.\ \cite{abr03}, and Hor\'ak \cite{hor04}).
One can watch the accuracy to which $\mathcal{E}$ is conserved in order
to verify the analytical approach against the exact numerical solution
(we show such comparison in  Figure~\ref{fig:qpo-ellipse} for the
pseudo-Newtonian case of Abramowicz et al.\ \cite{abr03}).

Equations (\ref{eq:nepc-32arh})--(\ref{eq:nepc-32ath}) can be merged in a
single equation by introducing the following parameterization:
\begin{equation}
 a_\rho^2=\xi^2 \mathcal{E},
 \quad
 a_\theta^2=\ratio{4}{9}\left(1-\xi^2\right)\mathcal{E}.
\end{equation}
Then the oscillations are described by two equations for $\xi(t)$ and
$\gamma(t)$,
\begin{eqnarray}
 \dot{\xi}\!\!&=&\!\!
 \ratio{1}{16}\beta\omega_\theta\xi^2\left(1-\xi^2\right)\mathcal{E}^{3/2}\sin\gamma,
 \label{eq:nepc-32xi}
 \\
 \dot{\gamma}\!\!&=&\!\!
 -\sigma\omega_\theta +\ratio{1}{4}\omega_\theta\mathcal{E}
 \left[\mu_r\xi^2 
 \right.\nonumber\\
 &&\left. + \ratio{4}{9}\mu_\theta\left(1-\xi^2\right) 
 +\ratio{1}{4}\beta\xi\left(3-5\xi^2\right)\mathcal{E}^{1/2}
 \cos\gamma\right],
 \label{eq:nepc-32gamma}
\end{eqnarray}
which satisfy the identity 
\begin{equation}
 \dot{\xi}\,\dd\gamma-\dot{\gamma}\,\dd\xi = 0.
\label{eq:df}
\end{equation}
Substituting for $\dot{\xi}$ and $\dot{\gamma}$ from 
eqs.~(\ref{eq:nepc-32xi})--(\ref{eq:nepc-32gamma}), eq.~(\ref{eq:df}) 
implies that
\begin{eqnarray}
 \mathcal{F} &\equiv & 8\left(1-\xi^2\right)\sigma+\mathcal{E}\left[\mu_r\xi^4-
 \ratio{4}{9}\mu_\theta\left(1-\xi^2\right)^2\right]
 \nonumber\\
 &&+\beta\mathcal{E}^{3/2}\xi^3\left(1-\xi^2\right)\,\cos\,\gamma
 \label{eq:nepc-32F}
\end{eqnarray}
is a second integral of motion. For a given value of energy $\mathcal{E}$, the 
system follows $\mathcal{F}=\const$ curves. In the other words, projection 
of the solution onto $(\gamma,\xi)$-plane satisfies
\begin{equation}
 \mathcal{F}(\gamma,\xi)=\const.
\end{equation}
This allows us to construct two-dimensional phase-space sections in
which the system evolution takes place. 

\begin{figure}[t]
\begin{center}
\includegraphics[width=0.5\textwidth]{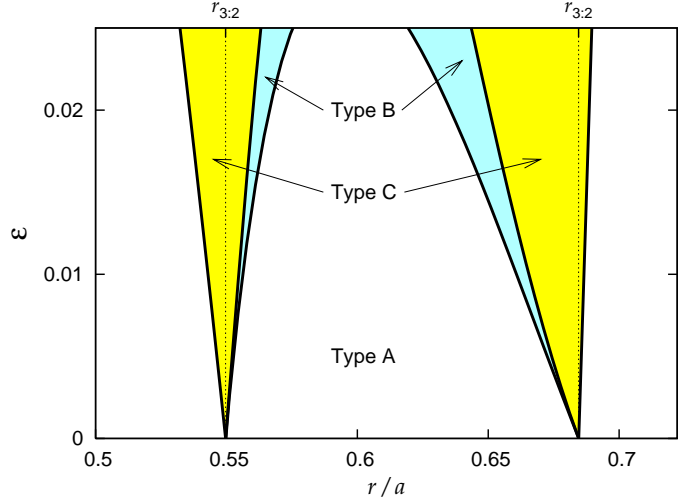}
\end{center}
\caption{The $3$:$2$ inner resonances in the gravitational field of a
spherical pseudo-Newtonian star and a ring ($a=9\rg$, $\mu=0.1$). The
regions of different phase-plane topology are identified in
$(r,\mathcal{E})$-plane. Three types can be distinguished according to
the number of critical points: A -- no critical point (the system is far
from resonance);  B -- one critical point; C -- two critical points.}
\label{fig:nepc-ring32}
\end{figure}

\subsection{Stationary points and the phase-plane topology}
Stationary points are given by the condition $\dot{\xi}=\dot{\gamma}=0$.
According to eq.~(\ref{eq:nepc-32xi}), $\gamma$-coordinate of these
points satisfies $\sin\gamma=0$, and therefore $\gamma=k\pi$ with $k$ 
being an integer. Substituting $\dot{\gamma}=0$ and $\cos\gamma=\pm1$ in
eq.~(\ref{eq:nepc-32gamma}), we find a cubic equation,
\begin{equation}
 -4\sigma+\left[\mu_r\xi^2+\ratio{4}{9}\mu_\theta\left(1-\xi^2\right)\right]\mathcal{E}
 \pm\beta\xi\left(3-5\xi^2\right)\mathcal{E}^{3/2}=0.
\label{eq:nepc-32steq}
\end{equation}
The solution determines $\xi$-coordinate of the stationary points. In
the case of small oscillations ($\mathcal{E}\ll1$), the solution can be
approximated by keeping only terms up to the linear one in
$\mathcal{E}$ in eq.~(\ref{eq:nepc-32steq}). We obtain
\begin{equation}
 \xi^2 = \frac{9\bar{\sigma}-\mu_\theta}{\ratio{9}{4}\mu_r-\mu_\theta},
\label{eq:nepc-32stsol}
\end{equation}
where $\bar{\sigma}\equiv\sigma/\mathcal{E}$. The first correction to
this solution is of the order of $\mathcal{E}^{1/2}$. Deviations
between $\xi$-coordinates of stationary points at odd and even 
multiples of $\pi$ are of the same order.

The solution
(\ref{eq:nepc-32stsol}) lies within the allowed range provided that
$\ratio{1}{4}\mu_r\lessgtr\bar{\sigma}\lessgtr\ratio{1}{9}\mu_\theta$ 
with the denominator $D\equiv\ratio{9}{4}\mu_r-\mu_\theta\lessgtr0$, 
respectively. This can be expressed in terms of energy
$\mathcal{E}$: given the detuning parameter $\sigma$, stationary points
appear in the $(\gamma,\xi)$ plane if the energy of oscillations
satisfies
\begin{equation}
 9\,\frac{\sigma}{\mu_\theta}\lessgtr\mathcal{E}\lessgtr4\,\frac{\sigma}{\mu_r}
 \quad\mathrm{for}\quad
 D\gtrless0.
\label{eq:nepc-32ineq}
\end{equation}

\begin{figure*}
\begin{center}
\includegraphics[width=0.325\textwidth]{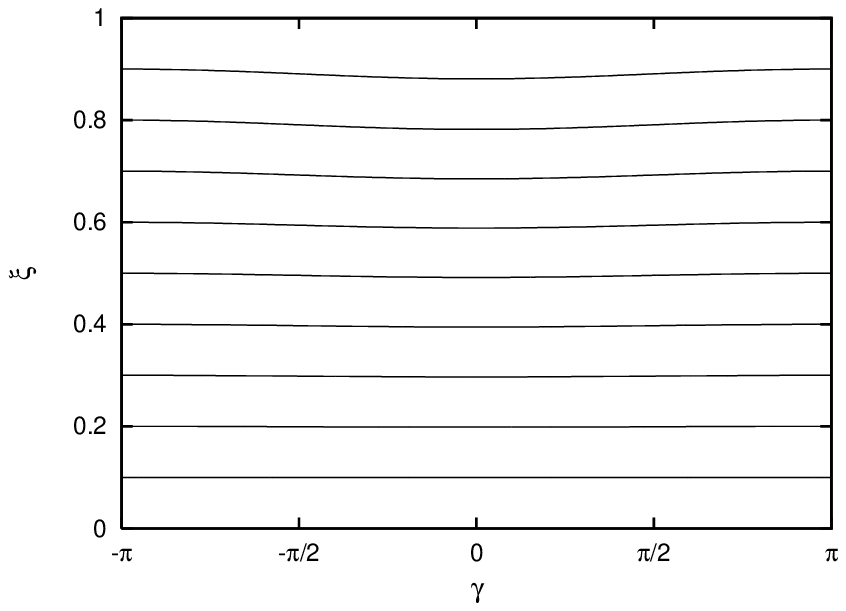}
\includegraphics[width=0.325\textwidth]{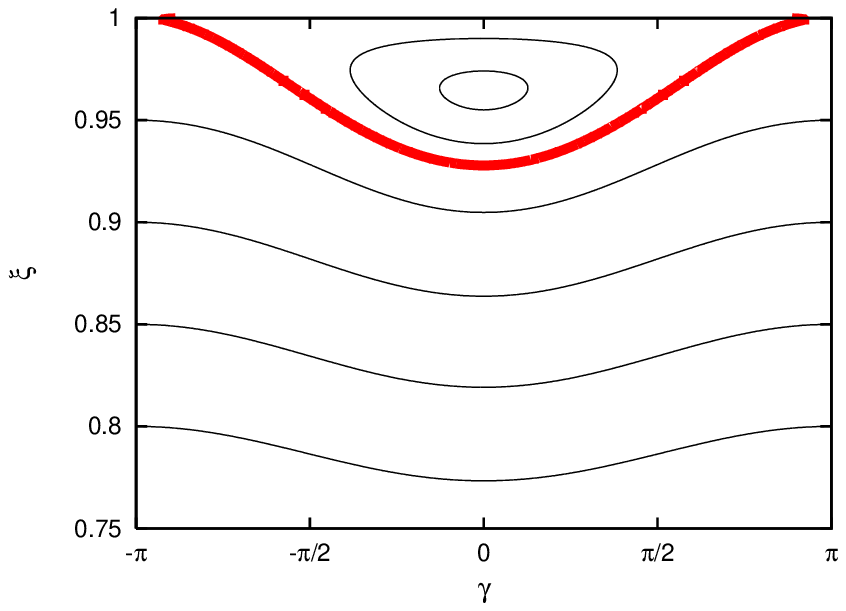}
\includegraphics[width=0.325\textwidth]{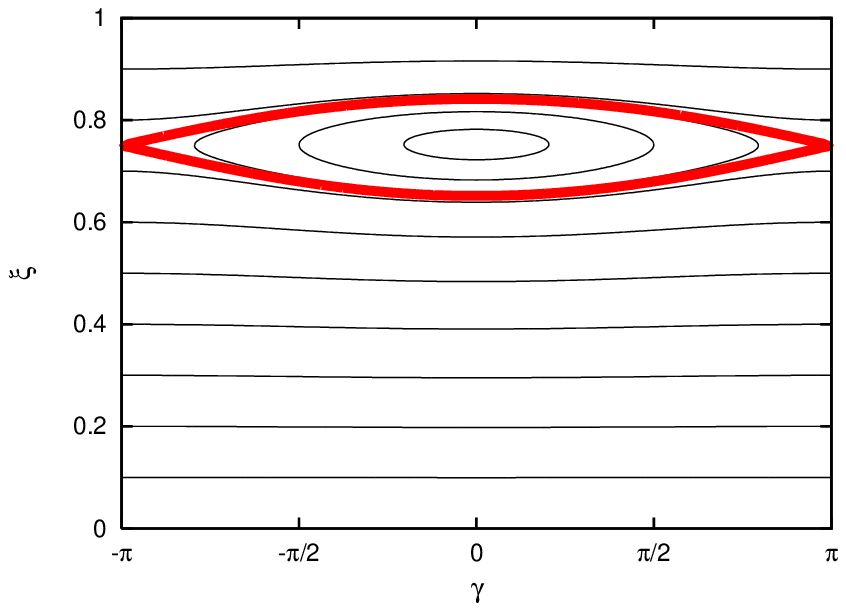}
\end{center}
\caption{Different topologies of the phase-space sections are shown in 
$(\xi,\gamma)$-plane. The system evolution follows the contour lines
$\mathcal{F}(\xi,\gamma)=\const$. Their shape determines a possible range
of the frequency change of the QPOs. Separatrix curve (thick line; 
cases B and C) divides the regions of circulating tracks from the regions 
of libration. Parameters and notation as in previous figure: A~(left; $r=0.6a$, $\mathcal{E}=0.02$), B~(middle; $r=0.65a$,
$\mathcal{E}=0.02$), C~(right; $r=0.55a$, $\mathcal{E}=0.05$).}
\label{fig:nepc-ring32a}
\end{figure*}

The examination of phase-plane topology near critical points leads to
the equation
\begin{equation}
 \left(\pder{\dot{\xi}}{\xi}-\lambda\right)\left(\pder{\dot{\gamma}}{\gamma}-\lambda\right)-
 \pder{\dot{\xi}}{\gamma}\,\pder{\dot{\gamma}}{\xi} = 0
\label{eq:nepc-11charp}
\end{equation}
for eigenvalues $\lambda$ of the system of linearized equations
(\ref{eq:nepc-32xi}) and (\ref{eq:nepc-32gamma}). Evaluating the partial
derivatives at the critical point and keeping only the terms of the
lowest order in $\mathcal{E}^{3/2}$, we obtain
\begin{equation}
 \lambda^2 = \pm\ratio{1}{72}\omega_\theta^2\beta\xi^3\left(1-\xi^2\right)D\,\mathcal{E}^{5/2}.
\end{equation}
Examining the sign of $\lambda^2$ demonstrates that the critical points of 
central topology alternate with those of saddle topology.

\subsection{The time dependence}
The equation for $\xi(t)$ can be derived by combining
eqs.~(\ref{eq:nepc-32xi}) and (\ref{eq:nepc-32F}). Eliminating
$\cos\gamma$, we arrive at the relation
\begin{equation}
 \mathcal{K}\dot{u}^2=F^2(u)-G^2(u),
\label{eq:nepc-32u}
\end{equation}
where we introduced a new variable $u(t)\equiv\xi^2$. The constant
$\mathcal{K}$ and functions $F(u)$ and $G(u)$ are defined by
\begin{eqnarray}
 \mathcal{K}&\equiv&
 \frac{1}{\mathcal{E}^{3/2}}\left(\frac{8}{\omega_\theta\beta}\right)^2,
 \\
 F(u) &\equiv&u^{3/2}(1-u)
 \\
 G(u) &\equiv&
 \frac{1}{\beta\mathcal{E}^{3/2}}\left[\mathcal{F}-8\sigma(1-u)-\mu_r\mathcal{E}u^2
 +\ratio{4}{9}\mu_\theta\mathcal{E}(1-u)^2\right].
\end{eqnarray}
The motion is allowed only for $\dot{u}^2\geq0$, and so the condition
$\pm F(u)=G(u)$ gives us two turning points, $u_1$ and $u_2$, between
which the evolutionary path oscillates. The functions $\pm F(u)$ and
$G(u)$ are plotted in Figure~\ref{fig:nepc-fg32}. 

The period of energy exchange can be found by integrating
eq.~(\ref{eq:nepc-32u}):
\begin{equation}
 T = \frac{16}{\beta \omega_\theta}\;\mathcal{E}^{-3/2} \int_{u_1}^{u_2} \frac{d u}{\sqrt{F^2(u) - G^2(u)}},
\end{equation}
where $T$ can be roughly approximated as
\begin{equation}
\label{eq:res32_T}
 T \sim \frac{16 \pi}{\beta \omega_\theta}\;\mathcal{E}^{-3/2}.
\end{equation}
Notice that this time-scale is longer than the period of individual
oscillations and it says how fast the system swaps itself between the
radial and vertical oscillation modes. The simple estimate
(\ref{eq:res32_T}) is quite precise in most parts of the phase space,
although it becomes inaccurate near stationary points, where the rate of
energy exchange slows down.    

Finally, we illustrate our results with a simple case, which we already
introduced at the beginning of the paper (sect.~\ref{sec:epc-ring}): the
gravitational field generated by a pseudo-Newtonian star and a narrow
circular ring. The resonant condition $\omega_\theta/\omega_r=3/2$ is
now fulfilled at three different radii, two of them lying between the
star and the ring, and the third one outside the ring. The resonances
occurring at first two radii are called the inner resonances, whereas
the latter is the outer resonance (not to be confused with `internal
resonance', which they are all). We restrict ourselves to the inner
resonances, for which we find $\sigma$, $\mu_r$, $\mu_\theta$ and
$\beta$ as functions of $r$. For a fixed radius, inequalities
(\ref{eq:nepc-32ineq}) give us the energy range of the oscillations. The
result is shown in Figure~\ref{fig:nepc-ring32}, where we identify three
different phase-plane topologies in the $(r,\mathcal{E})$-section. These
can be distinguished by the number of critical points and the shape of
separatrices. The topology change is evident in
Figure~\ref{fig:nepc-ring32a}, where two-dimensional plots are
constructed for the integral of motion $\mathcal{F}(\xi,\gamma)$.

\subsection{Frequencies of the resonant oscillations}
Equations (\ref{eq:D3a})--(\ref{eq:D3phi}) give the shift of actual
(observed) frequencies of oscillations, $\omega^\ast_r$ and
$\omega^\ast_\theta$, with respect to the eigenfrequencies $\omega_r$
and $\omega_\theta$:
\begin{equation}
 \omega^\ast_r = \omega_r + \dot{\phi}_r, \quad
 \omega^\ast_\theta = \omega_\theta + \dot{\phi}_\theta.
 \label{eq:qpos-corrections}
\end{equation}
These relations can be combined to find 
\begin{equation}
 2\omega_\theta^\ast-3\omega_r^\ast 
 = 2\omega_\theta-3\omega_r+2\dot{\phi}_\theta-3\dot{\phi}_r = \dot{\gamma}.
\label{eq:qpos-relation}
\end{equation}
The observed frequencies are in exact $3$:$2$ ratio if (and only if) the
time-derivative of the phase function $\gamma$ vanishes. An immediate
implication for the frequencies of stationary oscillations with constant
amplitudes is that they stay in exact $3$:$2$ ratio, even if the
eigenfrequencies may depart from it. Outside stationary points, it is
evident from Fig.~\ref{fig:nepc-ring32a} that $\dot{\gamma}=0$
represents turning points on libration tracks (those ones, which are 
encircled by the separatrix curve). Hence, eq.~(\ref{eq:qpos-relation})
discriminates between librating and circulating trajectories in the
$(\gamma,\xi)$-plane. Circulating trajectories span the full range of
$-\pi\leq\gamma<\pi$ and they do not contain any turning point;
$\dot{\gamma}$ remains nonzero and the twin frequencies never cross the
exact $3$:$2$ ratio in the region of circulation. On the other hand,
there are two points $\dot{\gamma}=0$ on each librating trajectory. In
such state the ratio of observed frequencies slowly fluctuates about
$3$:$2$.

\section{Conclusions}
We have discussed the resonance scheme for high-frequency QPOs via
multiple-scales analysis, assuming an axisymmetric conservative system
with two degrees of freedom. This approach provides a useful insight
into general properties that are common to different conceivable
mechanisms driving the oscillations, although it does not address the
question how the observed signal is actually formed and modulated.  In
our scenario, amplitudes and phases of the oscillations are mutually 
connected and they follow tracks in the phase space with distinct 
topologies. The particular form was assumed to couple the
oscillation modes via the non-spherical terms in the gravitational field
of a ring. We consider this to be a toy-model for rather general
behaviour, which should take place in any system governed by equations
of type (\ref{eq:ms-generalfrh})--(\ref{eq:ms-generalfth}).

We assumed the Newtonian (or the pseudo-Newtonian) description of the
central gravitational field with a perturbation by an aligned ring as an
example. The adopted form is not essential for general conclusions. In
fact, equations (\ref{eq:ms-generalfrh})--(\ref{eq:ms-generalfth}) cover
also the nearly-geodesic motion around a Schwarzschild black hole.
Compared with the pseudo-Newtonian case, general relativity does not
bring qualitatively new features, as long as the system is conservative
(additional terms will arise in the expansions, which then translate to
slightly different value of the resonance radius and to different 
duration of time intervals in physical units). A natural question arises
whether the gravitational field of a rotating black hole could provide
the perturbation required for the internal resonance in a surrounding
disc. We considered this possibility, however, it is  unlikely that Kerr
metric could by itself suffice: in the weak-field limit non-spherical
terms seem to be incapable of creating separatrices in the phase-space
sections discussed above, whereas in the full (exact, vacuum) Kerr
metric the special mathematical properties of the spacetime ensure the
integrability of the geodesic motion, and hence prevent the occurrence
of internal resonances. Therefore, the problem of a specific mechanism
launching and maintaining the oscillations remains unanswered.

Various options for the generalisation of our scheme could be motivated
by papers of other authors who proposed specific models including
non-gravitational forces (see P\'etri \cite{pet05} for a recent
exposition of the problem and for references). As a next step towards an
astrophysically realistic scheme, one should take dissipative and
non-potential forces into account, as well as non-axisymmetric
perturbations. These will allow our system to migrate across contours in
the phase-plane and to undergo transitions when crossing separatrices.
Such additional terms could also supplement the influence of
external forcing and initiate the oscillations of the system. The
internal resonance would then define the actual frequencies that are
excited; this way the  strong gravity unmasks itself.

\begin{acknowledgements}
VK appreciates fruitful discussions with participants of the Aspen
Center for Physics workshop `Revealing Black Holes', and both of us
thank for the hospitality of NORDITA (Copenhagen). We gratefully
acknowledge the financial support from the Czech Science Foundation
(refs.\ 205/06/P415 and 202/06/0041) and the Grant Agency of the Academy
of Sciences (ref.\ IAA300030510) that have been helping us at different
stages of the paper preparation. The Astronomical Institute has been
operated under the project AV0Z10030501.
\end{acknowledgements}

\end{document}